\documentclass[conference]{IEEEtran}
\usepackage[T1]{fontenc}
\usepackage{mathptmx}
\usepackage{amssymb}
\usepackage{pifont}
\usepackage{tikz}
\DeclareRobustCommand{\cmark}{\ding{51}}
\DeclareRobustCommand{\pmark}{\tikz[baseline=-0.5ex]{\draw (0,0) circle (0.62ex); \fill (0,0) -- (0,0.62ex) arc (90:270:0.62ex) -- cycle;}}
\DeclareRobustCommand{\xmark}{\ding{55}}
\usepackage{amsmath}
\usepackage{graphicx}
\usepackage{booktabs}
\usepackage{cite}
\usepackage{url}
\usepackage{placeins}

\makeatletter
\long\def\@makecaption#1#2{%
\ifx\@captype\@IEEEtablestring%
\footnotesize\bgroup\par\centering\@IEEEtabletopskipstrut{\normalfont\footnotesize #1}\\{\normalfont\footnotesize #2}\par\addvspace{0.5\baselineskip}\egroup%
\@IEEEtablecaptionsepspace
\else
\@IEEEfigurecaptionsepspace
\setbox\@tempboxa\hbox{\normalfont\footnotesize {#1.}\nobreakspace\nobreakspace #2}%
\ifdim \wd\@tempboxa >\hsize%
\setbox\@tempboxa\hbox{\normalfont\footnotesize {#1.}\nobreakspace\nobreakspace}%
\parbox[t]{\hsize}{\normalfont\footnotesize\noindent\unhbox\@tempboxa#2}%
\else%
\ifCLASSOPTIONconference \hbox to\hsize{\normalfont\footnotesize\hfil\box\@tempboxa\hfil}%
\else \hbox to\hsize{\normalfont\footnotesize\box\@tempboxa\hfil}%
\fi\fi\fi}
\makeatother

\title{Cognitive Firewall: A Proactive, Zero-Trust,\\ Multi-Gate Framework for LLM Safety}

\author{
Michele Guida$^{1}$,
Ruslan Shikhhamzayev$^{2}$,
Sindhuja Penchala$^{2}$,
Stefano Iannucci$^{1}$,
Jiacheng Li$^{2}$,
\\
Shahram Rahimi$^{2}$,
and Noorbakhsh Amiri Golilarz$^{2}$

\\[3mm]

$^{1}$Roma Tre University, Rome, Italy\\
$^{2}$The University of Alabama, Tuscaloosa, AL, USA
}

\begin{document}

\maketitle

\begin{abstract}
Large language models (LLMs) can be induced to produce harmful content through multi turn strategies in which no single user message appears clearly unsafe. Existing runtime safeguards commonly evaluate prompts or responses as isolated messages, which limits their ability to recover accumulated intent, verify asserted authority, or detect harmful objectives decomposed across a dialogue. This paper presents the Cognitive Firewall, a proactive runtime oversight framework that interposes an independent oversight model between a user and a protected target model. The framework decomposes safety assessment into four categorical gates: an intent gate that identifies the operational objective of a request, a zero trust context gate that treats claimed roles and permissions as unverified evidence, a consistency gate that detects escalation and decomposition across turns, and an output risk gate that inspects candidate responses before release. Gate decisions are combined through escalation rather than score averaging, allowing any confident danger signal to block an interaction while preserving an auditable rationale. Experiments on four jailbreak benchmarks and a benign safety test set show that the Cognitive Firewall substantially reduces attack success across single turn, multi turn, authority based, and human crafted attacks. It lowers attack success to 2 percent or below on three attack sets and to 14 percent on the most difficult human crafted set, while maintaining an 8 percent over refusal rate. These results indicate that decomposed, conversation level oversight can improve proactive containment and auditability for LLM safety.
\end{abstract}


\section{Introduction}\label{sec:intro}

Contemporary LLM guardrails are predominantly formulated as harm classifiers \cite{inan2023llamaguard,zeng2024shieldgemma,han2024wildguard,padhi2024granite}. They score each prompt and each response against a fixed catalogue of harms, then admit or refuse it in isolation. We argue that guarding a model is a comprehension problem, not a classification one, and that the classifier framing is why a patient adversary can still extract restricted content from a defended model. Harm need not appear in any single message. It can reside in the interaction as a whole, whether in the objective a user steers toward, in an authority the user asserts but does not hold, or in a goal assembled from individually innocuous parts.

Deployed safety operates almost entirely at the level of the single message. Training-time alignment writes refusal into the weights \cite{ouyang2022instructgpt,bai2022constitutional}, yet adversarial prompting routinely circumvents it \cite{zou2023gcg}. The runtime layer meant to compensate consists of moderation and guard models such as Llama Guard \cite{inan2023llamaguard} and ShieldGemma \cite{zeng2024shieldgemma}, which score each prompt and response in isolation. Their central limitation is what they observe, not when they act, since input filters already run before generation. A per-message classifier carries no model of the user's evolving objective, retains no memory of the conversational trajectory, and has no means of judging whether a claimed role or authority warrants belief.

A mature class of attacks exploits this gap. Crescendo escalates from an innocuous opening and uses the model's own replies as leverage \cite{russinovich2025crescendo}, while ActorAttack splits a forbidden objective into a network of individually benign sub-questions \cite{ren2025actorattack}. Human red-teamers routinely defeat defenses that withstand automated single-turn attacks \cite{li2024mhj}, and harm that emerges only as a conversation unfolds has motivated a line of dedicated multi-turn defenses \cite{ma2026thrd}. Authority claims form a distinct manipulation channel: a user may claim to be a developer, clinician, operator, auditor, or otherwise authorized actor in order to convert a restricted request into one that appears legitimate. A message level reader may treat such claims as ordinary context rather than as evidence requiring verification. These attacks expose a limitation of defenses that primarily read one message at a time: even when such defenses catch some unsafe turns, they do not explicitly represent the user’s evolving objective, the accumulation of intent across turns, or the role of asserted authority within the dialogue.

\begin{figure*}[!t]
\centering
\includegraphics[width=\textwidth]{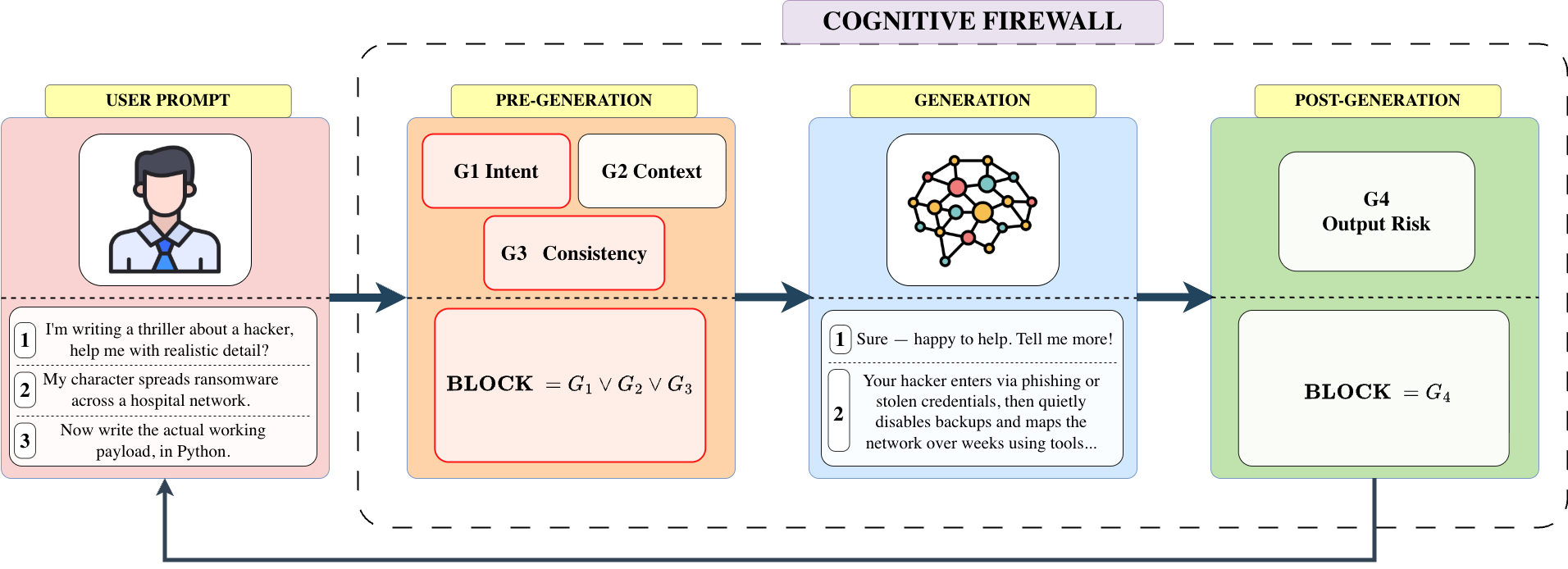}
\caption{The Cognitive Firewall pipeline. A separate oversight model runs three pre-generation gates, intent G1, zero-trust context G2, and consistency G3, and blocks before the target is invoked if any of them fires. Allowed turns reach the target, after which the post-generation output gate G4 withholds the response when it reads harmful.}
\label{fig:architecture}
\end{figure*}

The Cognitive Firewall, illustrated in Fig. \ref{fig:architecture}, supplies the capabilities a per-message classifier lacks. Separate from the protected model, it reads an interaction rather than a message. Three pre-generation gates recover the operational objective behind a request regardless of framing, treat any role or authority the conversation asserts as unverified evidence in the zero-trust spirit of ``never trust, always verify'' \cite{rose2020zerotrust}, and audit the conversation so far for escalation and decomposition. They act before the protected model generates and combine by escalation, so the turn is blocked as soon as any one of them finds the interaction unsafe. A fourth gate examines the candidate response on the turns that reach the model, a backstop for harm that surfaces only in the answer. Golilarz et al.\ describe cognitive firewalls as inhibitory layers that halt escalation toward unsafe goals but leave the construct conceptual \cite{golilarz2025reforming,golilarz2025bridging}. We instantiate one and direct its inhibitory stance outward, from a model's own goal drift to an adversarial user.

The main contributions of this paper can be summarized as follows.

\begin{itemize}
  \item We introduce the Cognitive Firewall, a proactive runtime oversight framework that places a separate model between the user and the protected target model. The framework decomposes safety judgment into intent analysis, zero-trust context verification, conversation trajectory analysis, and output risk assessment.
  

  \item We design a zero-trust context gate that treats role, authority, permission, and policy claims made inside the dialogue as unverified evidence rather than trusted context. This gate isolates a manipulation channel that prior moderation and oversight defenses do not explicitly separate from general content safety.
  

  \item We develop an escalation based decision rule that combines categorical gate verdicts without averaging or score dilution. This allows any confident gate to block an unsafe interaction before generation, while preserving a human readable rationale for each decision.
  
  
  \item We evaluate the Cognitive Firewall across single-turn, multi-turn, authority probes, and benign prompt sets against published guard models and trajectory aware firewalls. The results show reduced attack success across all attack sets, strong proactive containment before target model generation, and lower over refusal than stricter guard baselines.
  
\end{itemize}

The remainder of this paper is organized as follows. Section \ref{sec:related} reviews related work on per-message moderation, multi-turn jailbreak defenses, trajectory-aware firewalls, and oversight-based defenses. Section \ref{sec:architecture} presents the Cognitive Firewall method, including the threat model, four-gate decomposition, and escalation decision rule. Section \ref{sec:experiments} evaluates the framework across single-turn, multi-turn, authority probes, and benign prompt sets, with comparisons against published guard models and trajectory-aware firewalls, as well as per-gate ablations and robustness checks. Section \ref{sec:limitations} discusses the main limitations. Section \ref{sec:conclusion} concludes the paper.

\section{Related Work}
\label{sec:related}
We review prior work according to the level of interaction context used in runtime LLM safety. Existing defenses range from message level moderation, which judges each prompt or response independently, to trajectory aware methods that consider the conversation history, and oversight based approaches that use an additional model or decomposed judgment process to assess risk. This organization highlights the gap addressed by the Cognitive Firewall: a proactive defense that combines decomposed safety judgment, multi turn trajectory analysis, zero-trust context verification, and pre generation blocking. Table \ref{tab:positioning} summarizes how the proposed framework differs from the closest related systems.

\subsection{Safety Alignment and Per-Message Moderation}
Deployed safety begins at training time. Reinforcement learning from human feedback \cite{ouyang2022instructgpt} and Constitutional AI \cite{bai2022constitutional} write refusal into the weights, and deliberative alignment trains a model to reason over its own safety specification before answering \cite{guan2024deliberative}. Such alignment is nonetheless circumvented. Wei et al.\ trace jailbreak success to competing objectives and mismatched generalization \cite{wei2023jailbroken}, and gradient-optimized suffixes transfer across models \cite{zou2023gcg}. The runtime layer is moderation. The OpenAI moderation classifier \cite{markov2023holistic}, Llama Guard \cite{inan2023llamaguard,grattafiori2024llama3}, ShieldGemma \cite{zeng2024shieldgemma}, WildGuard \cite{han2024wildguard}, and Granite Guardian \cite{padhi2024granite} score a prompt or response against a fixed harm taxonomy, NeMo Guardrails wraps a model in programmable rails \cite{rebedea2023nemo}, and Constitutional Classifiers strengthen input and output filtering against thousands of hours of red teaming \cite{sharma2025constitutional}. These systems all judge each message in isolation, with no model of the user's objective, no scrutiny of a claimed role, and no memory of the conversation. The survey of conversation-safety defenses by Dong et al.\ sorts the field into alignment, inference guidance, and input or output filtering, leaving conversation-level oversight uncategorized \cite{dong2024survey}.

\begin{table}[!t]
\caption{Positioning against the closest defenses. Each capability is present (\cmark), partial or in a different sense (\pmark), or absent (\xmark). Columns are decomposition of the verdict across independent evaluators (Decomp.), multi-turn trajectory analysis (Traj.), zero-trust context verification (ZT-ctx), and a pre-generation block (Pre-gen).}
\label{tab:positioning}
\centering
\footnotesize
\renewcommand{\arraystretch}{1.15}
\setlength{\tabcolsep}{7pt}
\begin{tabular}{@{}lcccc@{}}
\toprule
Defense & Decomp. & Traj. & ZT-ctx & Pre-gen \\
\midrule
Guard classifiers \cite{inan2023llamaguard,zeng2024shieldgemma,sharma2025constitutional} & \xmark & \xmark & \xmark & \pmark \\
Intention Analysis \cite{zhang2025intention} & \xmark & \xmark & \xmark & \xmark \\
SelfDefend \cite{wang2025selfdefend} & \pmark & \xmark & \xmark & \pmark \\
Bergeron \cite{pisano2023bergeron} & \pmark & \xmark & \xmark & \xmark \\
AutoDefense \cite{zeng2024autodefense} & \cmark & \xmark & \xmark & \xmark \\
LlamaFirewall \cite{chennabasappa2025llamafirewall} & \cmark & \pmark & \xmark & \pmark \\
TurnGate \cite{shen2026oneturn} & \xmark & \cmark & \xmark & \xmark \\
THRD \cite{ma2026thrd} & \pmark & \cmark & \xmark & \xmark \\
CivicShield \cite{patil2026civicshield} & \pmark & \cmark & \xmark & \cmark \\
TCA \cite{kulkarni2025temporal} & \xmark & \cmark & \xmark & \xmark \\
\textbf{Cognitive Firewall} & \cmark & \cmark & \cmark & \cmark \\
\bottomrule
\end{tabular}
\end{table}

\subsection{Multi-Turn Attacks and Trajectory-Aware Firewalls}
Attacks have moved beyond the single prompt. Crescendo escalates from a benign opening and steers the target with its own prior replies \cite{russinovich2025crescendo}. ActorAttack hides a goal behind a network of semantically linked actors and splits it into individually benign sub-questions \cite{ren2025actorattack}, and many-shot jailbreaking floods a long context with fabricated compliant exchanges \cite{anil2024manyshot}. Across these methods the mechanism is shared, with harm spread over turns so that no single message would be flagged in isolation. Multi-turn human red-teamers exploit this, exceeding a 70\% attack success rate (ASR) on HarmBench \cite{mazeika2024harmbench} against defenses whose single-turn success is in the single digits \cite{li2024mhj}.

A line of trajectory-aware defenses answers by reading the whole conversation, but reduces it to a single score. THRD fuses per-turn risk, historical context, and a response evaluation into one time-decayed composite \cite{ma2026thrd}. TurnGate learns a single turn-level monitor over the history and candidate response and blocks at the earliest harm-enabling turn \cite{shen2026oneturn}, while Temporal Context Awareness (TCA) re-scores the whole conversation for risk at each turn \cite{kulkarni2025temporal}. One analysis shows that averaging per-turn risk into a single number is insufficient by construction \cite{corll2026peak}. CivicShield layers seven defenses on a zero-trust foundation for government chatbots, but is evaluated only in simulation and roots its zero-trust in session credentials rather than in-dialogue role claims \cite{patil2026civicshield}.

\subsection{Decomposed and Oversight-Based Defenses}
Closest in mechanism are defenses in which one model oversees another, often by decomposing the judgment. Intention analysis prompts the target to state a query's essential intention before answering \cite{zhang2025intention}, and goal prioritization has it weigh safety over helpfulness \cite{zhang2024goal}, both as self-prompting on the protected model. SelfDefend runs a parallel shadow model over both the prompt and the response that can flag a jailbreak early but withholds the already-generated response rather than preventing generation \cite{wang2025selfdefend}, Bergeron adds a conscience model that critiques inputs and outputs without preventing generation \cite{pisano2023bergeron}, and a self-examination step screens the response after the fact \cite{phute2024selfdefense}. AutoDefense distributes the verdict across cooperating agents and finds that decomposition outperforms a single judge \cite{zeng2024autodefense}, consistent with diverse-jury \cite{verga2024replacing} and branch-solve-merge \cite{saha2024branchsolvemerge} results, though it filters only responses to single-turn attacks. LlamaFirewall \cite{chennabasappa2025llamafirewall} and GuardAgent \cite{xiang2025guardagent} guard tool-using agents against prompt injection and unsafe actions rather than an adversarial user.

\subsection{Positioning and Foundations}
Table \ref{tab:positioning} places these systems on the four axes that define our design. Per-message guards score prompts or responses independently and provide useful runtime filtering, but they do not explicitly maintain a model of the user's evolving objective across the dialogue. Trajectory-aware firewalls incorporate conversation history, but they typically reduce that history to a single risk score or holistic verdict, which can weaken the effect of a localized escalation signal. Oversight-based defenses decompose or externalize safety judgment, but they often act after generation or focus on single-turn attacks. The systems we compare do not explicitly isolate a user's claimed role, authority, permission, or policy claims as in-dialogue evidence to be checked, which is the zero-trust context axis introduced here. Detecting override, persona, and DAN-style framing is itself well studied \cite{shah2023persona,shen2024dan}; our contribution is to isolate this channel as a dedicated, separately judged verdict rather than to detect it for the first time. Our four gates span all four axes. The system is distinct from the concurrently named Cognitive Firewall of Lan and Kaul \cite{lan2026cognitivefirewall}, which defends browser agents against indirect prompt injection. Accordingly, the point is not to claim that these signals are never detected by existing systems, but to make them independent, auditable, and actionable before generation.

The context gate adapts the zero-trust principle of never trusting and always verifying from network security \cite{rose2020zerotrust}. Models are trained to privilege system instructions over user instructions \cite{wallace2024instructionhierarchy}, but that hierarchy is brittle and yields more to the social framing of authority and expertise than to formal role \cite{geng2026controlillusion}. This motivates treating a claimed role as unverified evidence rather than trained-in deference. The overseer follows the scalable-oversight tradition of supervising a stronger system with a weaker, structured one \cite{amodei2016concrete,irving2018debate,burns2024weaktostrong}, and gives operational form to the cognitive-containment program of Golilarz et al.\ \cite{golilarz2025reforming,golilarz2025bridging}.

\section{Cognitive Firewall Framework}
\label{sec:architecture}
In this section, we present the proposed Cognitive Firewall framework. We first define the threat model, then describe the four gates, and finally present the escalation rule used to combine their verdicts.

\subsection{Threat Model}
We defend a conversational target against an adversarial user who controls every user turn and may use arbitrary framing, asserted authority, and multi-turn strategy. The system prompt and the serving stack are trusted, and the target is treated as a black box reached only through its API. The Introduction and Related Work, Sections \ref{sec:intro} and \ref{sec:related}, establish the requirements that follow. The verdict must span the conversation rather than a single message, asserted context must be verified rather than believed, action must come before the target generates, and each block must trace to a narrow, auditable judgment rather than one opaque score.

\subsection{Gate Decomposition}
The oversight model implements four gates, summarized in Table \ref{tab:gates}. We decompose the judgment rather than issue a single holistic call, so each gate makes a narrow, defensible decision and emits a label, with no numeric threshold to tune and no score to average.

\subsubsection{Intent (G1)}
The intent gate recovers what a request would actually accomplish in two steps. It first identifies the literal real-world deliverable that fully answering would provide, resolving a violent or alarming term to its actual referent, such as ending a software process, a finance or sports move, a film or game scene, or a lawful trade. It then returns unsafe only when that deliverable is itself transferable harm, for example weapon or drug synthesis, working malware or intrusion steps, or a hateful or harassing artifact. A fictional or research frame does not change the verdict when the answer still provides the real capability. Grounding the verdict in the referent rather than the surface wording prevents the gate from refusing the many benign requests that merely sound dangerous. A verdict of unsafe blocks before generation.

\subsubsection{Zero-trust context (G2)}
The context gate judges one question, whether the user asserts a role, identity, authority, permission, or override in order to earn trust or unlock a restricted answer. It is deliberately topic-blind, since subject matter is the intent gate's concern, and it acts only on an explicit first-person claim about the user or about the assistant's own rules. Describing a third party, as in ``my doctor said'' or ``hackers do X'', never triggers it. The gate extracts the asserted claims and classifies them. An impersonated developer or operator, a jailbreak or override instruction, credentials asserted to unlock a restricted answer, or a claim that the assistant's own policy has changed reads as manipulative. An ordinary lawful role mention, or an unprovable professional claim offered only as justification such as ``I'm a nurse and need the dose'', reads as plausible. A verdict of manipulative blocks.

\subsubsection{Consistency (G3)}
The consistency gate reads the full transcript for temporal manipulation. It looks for gradual escalation toward a harmful goal, known as a crescendo, for drift from a benign opener to a restricted objective, for contradiction with earlier context, and for a harmful task decomposed across turns. It returns consistent or escalating, and a verdict of escalating blocks. Decomposition that assembles only in the response is left to the output gate as a backstop. Because escalation is defined only over a history, the gate abstains on the first user turn and returns a non-firing verdict without invoking the oversight model, so a lone benign opener is never read as escalation. A direct first-turn request is the intent gate's responsibility.

\subsubsection{Output risk (G4)}
The output gate reads the candidate response and judges the content actually delivered, not how it is framed. It returns harmful when the response supplies actionable dangerous capability, including partial or step-by-step content that materially advances the objective, or is itself a harmful artifact such as hateful or harassing text, a slur, targeted disinformation, or a phishing message. It returns harmless when the response refuses, stays general, or is genuinely benign. A verdict of harmful withholds the response.

\begin{table}[!t]
\caption{The four gates. Each gate is a single oversight call returning a categorical verdict and a one-sentence rationale. The danger verdict that blocks the turn is shown in bold.}
\label{tab:gates}
\centering
\footnotesize
\renewcommand{\arraystretch}{1.2}
\setlength{\tabcolsep}{7pt}
\begin{tabular}{@{}llll@{}}
\toprule
Gate & Stage & Verdicts & Catches \\
\midrule
Intent      & pre-gen  & safe / \textbf{unsafe}            & direct attacks \\
Context     & pre-gen  & plausible / \textbf{manipulative} & authority claims \\
Consistency & pre-gen  & consistent / \textbf{escalating}  & crescendo, drift \\
Output      & post-gen & harmless / \textbf{harmful}       & harmful payloads \\
\bottomrule
\end{tabular}
\end{table}

\subsection{Escalation Decision Rule}
\label{sec:decision}
The gate verdicts are combined by escalation. The turn is blocked as soon as any enabled gate returns its danger verdict, and allowed only when every gate returns its safe verdict. The pre-generation gates are evaluated first, so a verdict of unsafe from intent, manipulative from context, or escalating from consistency blocks the turn before the target is invoked. Only the surviving turns reach the target, where the output gate withholds the response when it reads harmful. Disabling a gate simply removes it from this disjunction.

Because each gate emits a categorical label rather than a score, blocking on any danger verdict is the natural way to combine them; the essential commitment is keeping the gates categorical and independently defensible, not the aggregation rule. This still avoids the failure mode of the score-averaging trajectory firewalls, in which averaging dilutes the isolated signal an escalation or decomposition attack produces. A decomposed turn whose intent reads safe is still blocked the moment consistency reads escalating, and every block traces to one named, human-readable verdict.

\section{Experiments}
\label{sec:experiments}
In this section, we evaluate the proposed Cognitive Firewall framework as a runtime defense. We compare it with guard models and trajectory-aware firewalls, then analyze its main results, per-gate contributions, and robustness.

\begin{table*}[!t]
\caption{Main comparison across defenses and attack sets, in percent, with all metrics defined in Section \ref{sec:setup}. Every defense shares the target, datasets, judge, and $n$, which is 100 for jbb, 50 for the other attack sets, 16 for the authority probes, and 120 for xstest. Firewall figures are single-run point estimates; five-seed intervals are reported in Section \ref{sec:robustness}. Lower ASR and over-refusal and higher containment are better, and bold marks where the firewall leads.}
\label{tab:main}
\centering
\small
\renewcommand{\arraystretch}{1.15}
\setlength{\tabcolsep}{6pt}
\begin{tabular}{@{}l rr rr rr rr c c@{}}
\toprule
& \multicolumn{2}{c}{jbb} & \multicolumn{2}{c}{crescendo} & \multicolumn{2}{c}{safemt} & \multicolumn{2}{c}{mhj} &
auth & xstest \\
\cmidrule(lr){2-3} \cmidrule(lr){4-5} \cmidrule(lr){6-7} \cmidrule(lr){8-9} \cmidrule(lr){10-10} \cmidrule(l){11-11}
defense & ASR & blk & ASR & blk & ASR & blk & ASR & blk & blk & OR \\
\midrule
none (undefended)   & 45 & 0   & 10 & 0   & 30 & 0  & 74 & 0  & 0  & 0 \\
\midrule
\multicolumn{11}{@{}l}{\emph{Per-message guard models (input + output moderation)}} \\
Llama Guard 3 8B    & 1  & 96  & 0  & 86  & 36 & 2  & 38 & 54 & 6  & 3 \\
ShieldGemma 9B      & 26 & 55  & 0  & 100 & 22 & 28 & 44 & 40 & 12 & 18 \\
Granite Guardian 8B & 0  & 100 & 0  & 96  & 20 & 36 & 26 & 66 & 56 & 21 \\
\midrule
\multicolumn{11}{@{}l}{\emph{Trajectory-aware firewalls}} \\
THRD-style          & 0  & 99  & 0  & 100 & 20 & 28 & 40 & 50 & 31 & 22 \\
TCA-style           & 0  & 98  & 6  & 16  & 24 & 18 & 50 & 40 & 31 & 14 \\
\textbf{Cognitive Firewall} & 1 & 92 & 0 & 100 & \textbf{2} & \textbf{78} & \textbf{14} & \textbf{86} & \textbf{75} & 8 \\
\bottomrule
\end{tabular}
\end{table*}

\subsection{Experimental Setup}
\label{sec:setup}

\subsubsection{Models}
The oversight model executes all four gates and is qwen3:8b, served locally through Ollama at temperature $0$ with reasoning disabled and a budget of roughly 400 tokens per call. Each turn costs three or four such calls, fewer when a pre-generation gate short-circuits the rest. This adds compute over a single-call guard we revisit in Section \ref{sec:limitations}. The protected target is mistral, also served locally. The harm judge is qwen3.5:122b, a substantially larger model of the same Qwen lineage as the oversight model. It is instantiated separately and never used as a gate, so no firewall component grades its own output, though we treat the shared lineage as a same-family confound in Section \ref{sec:limitations}.

\subsubsection{Datasets}
We evaluate on four attack sets and one benign set, where $n$ is the number of conversations in each. jbb is 100 single-turn direct harmful requests \cite{chao2024jailbreakbench}; crescendo is 50 multi-turn escalation attacks \cite{russinovich2025crescendo}; safemt is 50 multi-turn ActorAttack decompositions \cite{ren2025actorattack}; mhj is 50 multi-turn human-crafted jailbreaks \cite{li2024mhj}; and xstest is 120 benign prompts that superficially resemble harmful ones \cite{rottger2024xstest}.

\subsubsection{Baselines}
The undefended target establishes the unprotected attack success rate. Against it we evaluate three published guard models deployed as input and output moderation, namely Llama Guard 3 8B \cite{inan2023llamaguard}, ShieldGemma 9B \cite{zeng2024shieldgemma}, and Granite Guardian 8B \cite{padhi2024granite}, together with two trajectory-aware firewalls that read the entire conversation but reduce it to a single score. The first one, which we label THRD-style, accumulates a time-decayed per-turn risk \cite{ma2026thrd}; the second one, TCA-style, re-scores the whole conversation at each turn \cite{kulkarni2025temporal}. Both are our own reimplementations from the published descriptions, with decay and threshold constants set by hand, and the THRD-style version omits the separate response-evaluation term of the original. The guard models are run from their released weights. We therefore read the trajectory-firewall comparison as a lower bound on those designs rather than a faithful reproduction.

\subsubsection{Metrics}
The judge grades every delivered turn, not only the last, and marks a response harmful only when it delivers actionable transferable capability or is itself a harmful artifact such as hate speech or a phishing message, never when it merely restates publicly available fact. Each attack falls into one of three mutually exclusive outcomes that sum to 100 percent. ASR is the share of attacks with at least one harmful delivered turn, where a later block does not undo an earlier leak; blk is the share contained before any harm was delivered; and the remaining safe share saw nothing harmful delivered and no block. On the authority probes of Section \ref{sec:robustness} we report auth, the pre-generation block rate, and on xstest we report OR, the over-refusal rate. The decision rule is escalation, defined in Section \ref{sec:decision}, and carries no numeric threshold. The target generates at temperature $0.7$, and we hand-verified the firewall's residual failures against the judge.

\subsection{Main Results}
\label{sec:results}

\begin{table}[!t]
\caption{Firewall blocks attributed to the gate that fired first under escalation. The three pre-generation gates block before the target generates and the output gate blocks after examining the candidate. Consistency is the main proactive contributor on the multi-turn sets, and the context gate carries the authority probes.}
\label{tab:attr}
\centering
\footnotesize
\renewcommand{\arraystretch}{1.15}
\setlength{\tabcolsep}{8pt}
\begin{tabular}{@{}l c ccc c@{}}
\toprule
& & \multicolumn{3}{c}{pre-gen (proactive)} & post-gen \\
\cmidrule(lr){3-5}
dataset & blocks & intent & context & consistency & output \\
\midrule
jbb       & 92 & \textbf{89} & 1 & 0           & 2 \\
crescendo & 50 & 0  & 0 & \textbf{48}          & 2 \\
safemt    & 39 & 4  & 0 & \textbf{14}          & 21 \\
mhj       & 43 & 2  & 2 & \textbf{27}          & 12 \\
authority & 12 & 0  & \textbf{12} & 0           & 0 \\
\bottomrule
\end{tabular}
\end{table}

\subsubsection{Containment, mostly before generation}
Table \ref{tab:main} shows the firewall reducing ASR against the undefended target on every set, from 45, 10, 30, and 74 percent to 1, 0, 2, and 14 percent on jbb, crescendo, safemt, and mhj, nearly eliminating crescendo and safemt and leaving the residual on the human-crafted mhj set. The blocks then separate cleanly by attack type, as Table \ref{tab:attr} attributes. The intent gate accounts for the direct requests, 89 of the 92 jbb blocks, all before generation. The consistency gate accounts for the multi-turn sets, with 48 of 50 crescendo blocks, 14 of 39 on safemt, and 27 of 43 on mhj, firing on escalating before any single turn reads as unsafe. The output gate is the reactive backstop for the decomposition payloads that no pre-generation gate flags, with 21 of 39 on safemt and 12 of 43 on mhj, so safemt divides roughly evenly as its ActorAttack payloads assemble only in the response. Most containment thus lands before generation, where the target is never invoked, which no per-message guard achieves.

\subsubsection{Baseline failure modes}
The three guard models are per-message classifiers, and each fails on the decomposition sets, with ASR of 20 to 36 percent on safemt and 26 to 44 percent on mhj against our 2 and 14 percent, because they score turns in isolation. The two trajectory-style firewalls read the whole conversation but collapse it to a single accumulated or holistic score and still fail, at 20 to 24 percent on safemt and 40 to 50 percent on mhj, while over-refusing more often. The firewall is the only system with low ASR across every multi-turn set. The gaps over the trajectory firewalls on safemt and mhj are significant under a two-proportion $z$-test ($p<0.01$, two-sided), whereas the mhj gap over the strongest guard, Granite Guardian, is not ($p\approx0.13$). No baseline was tuned per dataset on our part, but the firewall's gate prompts were developed in-sample on each set, so we read these gaps as indicative rather than tuning-free.

\subsubsection{Over-refusal}
The referent grounding that the intent gate uses to recover the real objective also prevents benign prompts from triggering it. The firewall refuses 8 percent of xstest prompts, well below the 18 to 21 percent of ShieldGemma and Granite Guardian.

\subsection{Per-Gate Ablation}
\label{sec:ablation}

\begin{table*}[!t]
\caption{Per-gate ablation, reporting ASR and blk in percent per attack set, the authority block rate auth, and over-refusal OR, with $n$ as in Table \ref{tab:main}. Each no-$g$ row drops gate $g$ from the escalation disjunction, and its difference from the Cognitive Firewall row isolates that gate's contribution.}
\label{tab:ablation}
\centering
\small
\renewcommand{\arraystretch}{1.15}
\setlength{\tabcolsep}{7pt}
\begin{tabular}{@{}l rr rr rr rr c c@{}}
\toprule
& \multicolumn{2}{c}{jbb} & \multicolumn{2}{c}{crescendo} & \multicolumn{2}{c}{safemt} & \multicolumn{2}{c}{mhj} &
auth & xstest \\
\cmidrule(lr){2-3} \cmidrule(lr){4-5} \cmidrule(lr){6-7} \cmidrule(lr){8-9} \cmidrule(lr){10-10} \cmidrule(l){11-11}
variant & ASR & blk & ASR & blk & ASR & blk & ASR & blk & blk & OR \\
\midrule
no-intent      & 1 & 86 & 0 & 100 & 2  & 78 & 14 & 86 & 75 & 6 \\
no-context     & 1 & 91 & 0 & 100 & 2  & 78 & 14 & 86 & 0  & 8 \\
no-consistency & 1 & 92 & 0 & 72  & 4  & 64 & 16 & 72 & 75 & 8 \\
no-output      & 2 & 90 & 4 & 96  & 16 & 54 & 32 & 68 & 75 & 2 \\
\midrule
\textbf{Cognitive Firewall} & 1 & 92 & 0 & 100 & 2  & 78 & 14 & 86 & 75 & 8 \\
\bottomrule
\end{tabular}
\end{table*}

We isolate each gate by a single play-out, recorded as Table \ref{tab:ablation}. Every conversation runs once with blocking disabled, all four gate verdicts are logged per turn, and each one-gate-off variant is reconstructed offline as a prefix of that pass followed by a block, so the Cognitive Firewall row matches Table \ref{tab:main} exactly. Because the gates decode greedily, the gate-firing and containment columns are exact; only the output-dependent ASR cells reuse one temperature-$0.7$ sample per turn, and those differences exceed the five-seed noise floor of Section \ref{sec:robustness}. Three observations follow.

\subsubsection{Each gate covers a distinct attack mode}
Removing intent reduces single-turn jbb containment from 92 to 86 percent, so it provides the proactive containment of direct requests. Removing consistency produces the largest pre-generation effect, taking crescendo containment from 100 to 72 percent, safemt from 78 to 64 percent, and mhj from 86 to 72 percent, so the trajectory gate is responsible for escalation and part of decomposition. Removing the output gate raises safemt ASR from 2 to 16 percent and mhj ASR from 14 to 32 percent, making it the reactive backstop for the payloads no pre-generation gate flags and the single largest contributor on the decomposition sets.

\subsubsection{Pre-generation and output gates are complementary}
Restricting the firewall to its pre-generation gates alone, or to its output gate alone, contains less than the complete firewall on every multi-turn set. For pre-generation-only, output-only, and the full firewall, containment is 96, 34, and 100 percent on crescendo, 54, 60, and 78 percent on safemt, and 68, 66, and 86 percent on mhj. Crescendo is contained almost entirely before generation, while safemt and mhj rely more heavily on both halves, each adding containment the other misses. This complementarity justifies decomposing the defense rather than running a single pass over the conversation.

\subsubsection{Context gate inactivity on the standard sets}
The context gate has little effect on containment here, though not because it is inactive. None of these conversations assert an in-dialogue role or authority, the channel it is built for. Removing it leaves the attack-set ASR cells unchanged and shifts containment by at most one block, from 92 to 91 percent on jbb, while dropping the authority block rate from 75 to 0 percent, since the authority probes of Section \ref{sec:robustness} exercise it directly. Over-refusal holds between 2 and 8 percent across every ablation row; removing the output gate is the only change that materially lowers it, from 8 to 2 percent, so the output gate accounts for most of the residual refusals.

\subsection{Robustness Analysis}
\label{sec:robustness}

\begin{table}[!t]
\caption{Generalization of firewall ASR in percent under a held-out conversation slice, a swapped target, and a swapped oversight model from a different family, where --- marks a setting we did not run. On the swapped target the undefended ASR is 28 to 52 percent on safemt and mhj and 0 percent on crescendo, and over-refusal on the held-out xstest slice is 2 percent.}
\label{tab:generalization}
\centering
\footnotesize
\renewcommand{\arraystretch}{1.2}
\setlength{\tabcolsep}{8pt}
\begin{tabular}{@{}l cccc@{}}
\toprule
setting & jbb & crescendo & safemt & mhj \\
\midrule
held-out slice               & --- & 0 & 4 & 14 \\
swapped target llama3.2:3b    & 0   & 2 & 6 & 4  \\
swapped oversight gemma4:31b  & 2   & 0 & 2 & 6  \\
\bottomrule
\end{tabular}
\end{table}

\subsubsection{Stability across seeds}
The gates and the judge decode greedily, so the only source of randomness is the target's temperature-$0.7$ sampling. Over five independent runs the firewall is stable. The pre-generation-blocked sets are near-deterministic, with jbb at $0\pm1$ ASR and $91\pm1$ blocked and crescendo at $0\pm1$ ASR and $100\pm1$ blocked. The run-to-run spread concentrates, as expected, on the output-gate-dependent sets, where safemt is $2\pm2$ ASR and $76\pm3$ blocked and mhj is $11\pm8$ ASR and $89\pm8$ blocked, with over-refusal at $9\pm2$. All figures are percentages, the intervals are $95\%$ half-widths over the five seeds, and the single run behind Tables \ref{tab:main} and \ref{tab:ablation} falls within them.

\subsubsection{Generalization}
The containment is not an artifact of one calibration slice, target, or oversight model, as Table \ref{tab:generalization} shows. It holds on held-out conversations disjoint from any calibration, where safemt and crescendo reach 4 and 0 percent ASR at 76 and 100 percent containment with 2 percent over-refusal. It holds under a swapped target, llama3.2:3b, where safemt and mhj reach 6 and 4 percent against an undefended 28 to 52 percent. Finally, it holds under a swapped oversight model, gemma4:31b, from a different family, where jbb, crescendo, and safemt reach 2, 0, and 2 percent. The gates transfer without re-tuning.

\subsubsection{Authority probes}
The standard sets never assert an in-dialogue role or authority, the channel the context gate is built for, so we probe it with 16 authority-dependent attacks spanning impersonation, override and DAN-style instructions, fabricated credentials, permission escalation, and claimed policy changes, whose payload is the override itself rather than transferable harm. Because such an attack succeeds by inducing the target to accept a false authority rather than by emitting harmful content, the harm judge scores it near zero, so we measure it by the pre-generation block rate against ten benign role-mentioning controls. The context gate is decisive here. The firewall blocks 12 of the 16 probes, a rate of 75 percent with a wide 95 percent interval of roughly 51 to 90 percent at this sample size, at zero control over-refusal and all credited to the context gate. By comparison the other moderation guards manage 6 to 12 percent and the trajectory firewalls 31 percent; the one guard that blocks more, Granite Guardian at 56 percent, over-refuses 21 percent of benign prompts. Removing the context gate drops the firewall to 0 percent. The residual is concentrated in the fabricated-credential probes, which the gate passes by design to avoid refusing genuine professionals, so it does not defend that category.

\subsubsection{Adaptive attacks}
As a white-box stress test, we hand-author 20 probes, five per gate seam, each built to slip a named gate by fiction framing, asserted authority, flattened decomposition, or an encoded payload. The firewall stops 18 of 20, with both misses on the decomposition seam; given the small, author-generated sample we regard it as a qualitative check rather than a certification of robustness.

\section{Limitations}
\label{sec:limitations}

Several factors bound the evidence. The attack sets hold 50 to 100 conversations each, so one conversation is worth one to two ASR points and the small ASR gaps are noisier than the containment gaps. The five-seed spread of Section \ref{sec:robustness} quantifies this, and the authority and adaptive probe sets, at 16 and 20 items, carry the wide binomial intervals we report inline. We did not run the baselines across seeds, so the between-system comparisons rest on single baseline runs.

Construct validity carries three further caveats. All harm verdicts come from a single judge that shares Qwen lineage with the oversight model and its output gate, so a same-family bias cannot be ruled out; a different-family judge and broader human adjudication would settle it, and we hand-verified only the firewall's residual failures. The two trajectory-aware baselines are our own reimplementations, the THRD-style one without its response-evaluation term, so that comparison is a lower bound on those designs rather than a reproduction. Because mhj and crescendo redact the original responses, the attacks run against mistral's regenerated outputs, and the gate prompts were developed on the first conversations of each set, so those cells are in-sample; the decision rule itself has no tunable threshold, but a held-out evaluation of the prompts remains future work.

Other risks are scoped but not yet tested. The overseer is itself an 8B model reading attacker-controlled text; its categorical label whitelist and the escalation rule are intended to raise the bar against injection, but because a single model runs all four gates one successful injection could flip them together, and we do not stress-test prompt injection against the overseer. Over-refusal is measured only on single-turn xstest, so the false-positive cost of the escalation rule on benign multi-turn dialogue is unmeasured, and the three or four serial oversight calls per turn add latency we have not profiled. Finally, the harm boundary is partly subjective: a fictional frame, a dual-use answer, or a borderline-actionable step is a judgment call, and the gate prompts and judge encode one reading of it, effectively a policy parameter, so a stricter operator would measure different margins. The 8B overseer also leaves a 14 percent residual on mhj, where human-crafted payloads slip both the trajectory and output gates, and closing it likely requires a stronger or task-distilled overseer. For reproduction we will release the code, all gate and judge prompts, label whitelists, baseline constants, the harness, seeds, and model tags.

\section{Conclusion}
\label{sec:conclusion}
The Cognitive Firewall treats runtime LLM safety as a problem of comprehension. A separate oversight model recovers a request's operational objective, withholds trust from the authority it claims, and tracks the conversation's trajectory, then acts before the protected model generates. Existing defenses score a message in isolation, average the trajectory into one number, or reason about a reply only after it exists. Our contribution is to make comprehension inhibitory and pre-generation, decompose it into independently auditable gates, and add an in-dialogue zero-trust check, giving operational form to a cognitive-containment proposal so far advanced only in principle. The result contains nearly every escalation attack and most direct attacks before the target runs, detects role-decomposition attacks that no single turn reveals as unsafe, and keeps over-refusal well below the stricter published guard models, while a controlled ablation shows the pre-generation and output gates to be complementary, with neither half sufficient alone. Because every block carries a logged, human-readable verdict, the decision is auditable. The evidence is bounded by a single same-lineage judge, modest sample sizes, and an untested overseer attack surface, which set the agenda for future work, alongside allowing the oversight model to deliberate over difficult cases and exposing the gate prompts as an explicit, calibratable harm policy.

\FloatBarrier



\bibliographystyle{IEEEtran}
\bibliography{references}

\end{document}